# Photoemission Insights to Electronic Orders in Kagome Superconductor $A$V$_3$Sb$_5$

Yigui Zhong[1*], Jia-Xin Yin[2*], Kosuke Nakayama[3*]

[1]*Insitute for Solid States Physics, The University of Tokyo, Kashiwa 270-8581, Japan*
[2]*Department of Physics, Southern University of Science and Technology, Shenzhen 518055, China*
[3]*Department of Physics, Graduate School of Science, Tohoku University, Sendai 980-8578, Japan*

Kagome superconductors $A$V$_3$Sb$_5$ ($A$ = K, Rb and Cs) have attracted considerable attention due to their intriguing combination of unique electron correlations and nontrivial band topology. The interplay of these fundamental aspects gives rise to a diverse array of exotic electronic phenomena, including superconductivity and charge density wave (CDW) states. In this review, we present recent advancements in the study of the electronic band structure of $A$V$_3$Sb$_5$ using angle-resolved photoemission spectroscopy (ARPES), including the identification of the multiple van Hove singularities near the Fermi level and their close relationship with the CDW transition, spectroscopic features related to CDW-induced symmetry breakings, as well as direct observations of nodeless superconducting gaps and moderate electron-phonon couplings through ultrahigh-resolution ARPES, providing critical insights into the origins of CDW order and electron pairing symmetry. By synthesizing these key ARPES findings, this review aims to deepen our understanding of kagome-related physics.

*Corresponding authors: yigui-zhong@issp.u-tokyo.ac.jp; yinjx@sustech.edu.cn; k.nakayama@arpes.phys.tohoku.ac.jp

## 1. Introduction

The exploration of quantum materials that intertwine electron correlations and band topology represents an exciting frontier in condensed matter physics. This convergence of electron correlations, pivotal for unconventional superconductivity [1,2], and band topology, essential for realizing nontrivial topological states [3,4], opens avenues for uncovering intriguing quantum phenomena. Particularly noteworthy are the transition-metal based kagome metals [5-8], which serve as exemplary platforms showcasing this convergence. The unique geometry of the kagome lattice with three sublattices [Fig. 1(a)] leads to the cancellation of the electronic wavefunctions, giving rise to a topological flat band structure. Additionally, owing to their analogous group symmetry with graphene, these kagome metals exhibit a Dirac band and van Hove singularities (VHSs) in their electronic band structures [Fig. 1(b)]. Thus, kagome metals serve as a fertile platform for investigating quantum phenomena arising from the intricate interplay between electron correlations and band topology. Recent attention has been drawn to non-magnetic $A$V$_3$Sb$_5$ compounds [9,10] [Figs. 1(c) and 1 (d)], in which a diverse array of electronic phenomena has been unveiled. These include charge density wave (CDW) [11-16], superconductivity [17-19], nematic order [11,12,20,21], pair density wave [17] and the giant anomalous Hall effect [22,23]. Moreover, the CDW order, characterized by its three-dimensional reconstruction, exhibits exotic time-reversal symmetry (TRS) breaking [14,15] and rotational symmetry breaking [12,13]. Through the application of pressure to suppress the CDW, intriguing interplays between CDW and superconductivity emerge, manifesting in the appearance of two superconducting domes [24,25].

In this paper, we undertake a comprehensive review of the research progress concerning CDW and superconductivity in $A$V$_3$Sb$_5$ kagome metals, with a particular focus on insights garnered through angle-resolved photoemission spectroscopy (ARPES). We scrutinize the band structure in the normal state, highlighting the presence of the multiple VHSs. Subsequently, we scrutinize the CDW gap and the chemical substitution effects on the CDW order, elucidating the intimate relationship between VHSs and the CDW transition. Furthermore, we present spectroscopic evidence of the symmetry breakings in the CDW state and discuss their potential relationship with CDW. Finally, we review the ARPES findings concerning electron-phonon coupling and superconducting gap symmetry, offering our understandings on the superconducting mechanisms underlying kagome superconductors.

## 2. Multiple van Hove singularities in the band structure

Among various kagome metals, $A$V$_3$Sb$_5$ stands out as a rare example predicted to host

VHSs near the Fermi level ($E_F$), thereby sparking interest in the relationship between VHSs and exotic physical properties [26-36]. In the tight-binding model for the kagome lattice [Fig. 1(b)], two VHSs appear at the M point of the Brillouin zone at electron fillings of $n$ = 5/12 and 3/12. The former is termed a p-type VHS as it originates purely from a single sublattice [37] [A, B, or C in Fig. 1(a)], as visible from the sublattice-resolved Fermi surface in Fig. 1(e), while the latter is termed an m-type VHS because of its mixed sublattice character [AB, BC, or CA; see Fig. 1(f)]. In $A$V$_3$Sb$_5$, density-functional theory (DFT) calculations predict the formation of multiple p- and m-type VHSs (VHS1-VHS4) at the M point due to multi-orbital characteristic in the V kagome lattice [29,30,38,39] [Fig. 1(g)]. VHS1 and VHS2 are p-type, with the $d_{x^2-y^2}$ and $d_{yz}$ orbital characters, respectively [the orientations of the $x$ and $y$ axes are shown in Fig. 1(d); the $d_{x^2-y^2}$ orbital forming VHS1 is also illustrated as an example]. On the other hand, VHS3 and VHS4 are m-type, with the former having a $d_{xy}$ orbital character and the latter a $d_{xz}$ orbital character. In addition to VHSs, Dirac-cone dispersions are formed at the K point by the kagome lattice symmetry.

The first ARPES results were reported simultaneously with the discovery of superconductivity in CsV$_3$Sb$_5$ by Ortiz *et al* [10]. The observed Fermi surface consists of circular and hexagonal pockets at the Brillouin zone center and corner, respectively [Fig. 1(h)]. A comparison with the DFT calculations suggests that the former is of Sb $p$ orbital origin, and the latter is derived from V $d$ orbitals with the hexagonal shape supporting an electron filling near the VHS point. Further ARPES measurements by Hu *et al*. [39] and Kang *et al*. [38] revealed that VHS1 and VHS4 are close to $E_F$, and VHS2 is slightly below $E_F$ [Fig. 1(i)]. A detailed band dispersion analysis further demonstrates that VHS1 is an unconventional higher-order VHS characterized by $k^4$ band dispersion along the MK direction [38,39] [Figs. 1(j)-1(l)]. The flattened band dispersion would lead to a strong power-law divergence of the density of states, enhancing instability toward ordered phases [40,41]. These ARPES observations provide an important starting point for understanding the intriguing electronic properties of $A$V$_3$Sb$_5$.

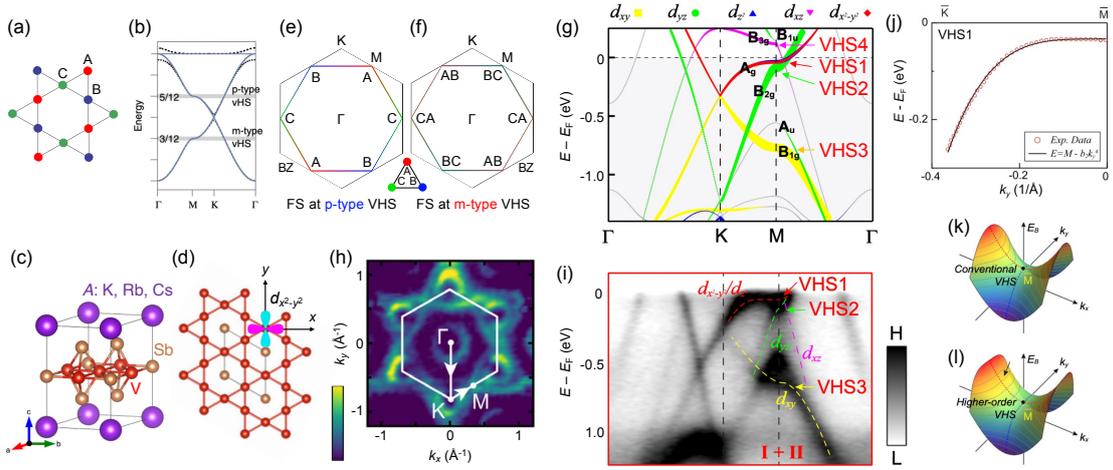

**Fig. 1.** (Color online) (a) Schematic of kagome lattice. Atoms in three different sublattices A-C are displayed in different color codes. (b) Typical band structure of the kagome lattice obtained by a tight-binding model with and without spin-orbit coupling [38] (solid and dashed lines, respectively). Two VHSs at electron fillings of $n$ = 5/12 and 3/12 are marked with grey shades. (c) and (d) Crystal structure of $A$V$_3$Sb$_5$ [9]. (e) Fermi surface of the kagome lattice at the $n$ = 5/12 filling with p-type VHS at $E_F$. (f) Same as (e) but for the $n$ = 3/12 filling with m-type VHS at $E_F$. Red, blue, and green colors of the Fermi surface indicate the contribution from A, B, and C sublattices, respectively. (g) Orbital-resolved DFT calculations [39] for CsV$_3$Sb$_5$. (h) ARPES intensity plot at $E_F$ measured at $T$ = 80 K in CsV$_3$Sb$_5$ [10]. (i) ARPES intensity measured along the ΓKMΓ high-symmetry cut in CsV$_3$Sb$_5$ [39]. Experimentally observed VHSs are marked by arrows. (j) Experimental band dispersion of VHS1 along the MK direction (red circles), together with the numerical fitting result using $k^4$ function (solid curve). (k) and (l) Schematics of conventional VHS and higher-order VHS, respectively. In (l), the band dispersion is flatter along the $k_y$ axis (MK direction) due to the $k^4$ band dispersion relationship seen in (j). This figure is reproduced from Ref. [9] (© 2019 American Physical Society), Ref. [10] (© 2020 American Physical Society), Refs. [38] and [39] (© 2022 The Authors).

## 3. Close relationship between VHSs and CDW transition
### 3.1 CDW gap anisotropy

The CDW state of $A$V$_3$Sb$_5$ forms a 2×2 in-plane superlattice [11,12,17,42-44], presumably due to star-of-David (SoD) or tri-hexagonal (TrH) distortion of V atoms [26] [Figs. 2(a) and 2(b)]. High-resolution ARPES measurements uncovered a close relationship of this 2×2 CDW with VHS. For instance, Nakayama *et al.* [16] reported that the hole branch of the VHS1 band, which reaches $E_F$ on the ΓM cut in the normal state [Figs. 2(c) and 2(e)], exhibits an energy gap (CDW gap) below CDW transition temperature ($T_{CDW}$) [black dashed line in Fig. 2(d) and red circles in Fig. 2(f)]. The large gap size of approximately 70 meV categorizes $CsV_3Sb_5$ as a strong-coupling CDW material. The momentum-dependent study shows that the CDW gap size on the V $d_{x^2-y^2}$ Fermi surface [red color in Fig. 2(g)] rapidly decreases away from VHS1 around the M point (Fermi surface angle $\theta \sim 0$ deg) and becomes negligibly small on the ΓK line ($\theta = 30$ deg), as plotted by red circles in Fig. 2(h). Similar gap anisotropy is observed along the V $d_{yz}$ Fermi surface [blue color in Fig. 2(g)] but with a much smaller maximum gap size [approximately 20 meV; see blue circles in Fig. 2(h)]. Furthermore, no CDW gap is observed on the Sb $p$ Fermi surface [green color in Figs. 2(g) and 2(h)]. Similar results have been reported for $KV_3Sb_5$ and $RbV_3Sb_5$ [45,46]. The largest CDW gap opening near VHS1 is consistent with a theoretical proposal that electron scattering via $q_1$-$q_3$, connecting VHS1 at different M points, significantly contributes to stabilizing $3q$ CDW. In this scattering process, the p-type nature of VHS1 is thought to be the key factor. Specifically, the p-type VHS suppresses the scattering channel through on-site Coulomb interactions because electron wave functions at the VHS1 points are localized on different sublattices [37] [Fig. 1(e)]. This sublattice interference effectively enhances the importance of nonlocal (i.e., nearest-neighbor) Coulomb interactions and potentially leads to nearest-neighbor bond density wave order [28,29,36,47,48], compatible with SoD and TrH CDW distortions. It is worth noting that the p-type VHS filling is also predicted to promote quantum interference between paramagnons, and the triple-$q$ CDW state can also be stabilized by this paramagnon interference effect [33].

In the CDW phase, an additional band appears at ~20 meV [purple triangles in Fig. 2(f)]. It was initially suggested to originate from the CDW-gapped state at $k_z = \pi$, while an alternative interpretation proposed it as the quasiparticle peak of VHS1 at $k_z = 0$, with the ~70 meV peak being a hump structure indicating strong electron-phonon coupling [49]. Subsequent studies supported that the 20-meV peak results from the band at $k_z = \pi$ due to intrinsic out-of-plane band folding from the 3D CDW [50-53], as detailed in the next section. The large CDW gap in VHS1 (~70 meV) agrees with that reported by optical measurements [54], and the smaller CDW gaps (~20 meV) at $k_z = \pi$ and other Fermi surface segments at $k_z = 0$ are comparable to the gap size reported by scanning tunneling microscopy (STM) measurements [11,55,56], highlighting multiple energy scales in $CsV_3Sb_5$.

It is also remarked that VHS4 is located near $E_F$, similar to VHS1, as mentioned in the previous section. According to the work by Kang *et al.* [38], the $d_{xz}$ band forming VHS4 indeed produces a hexagonal Fermi surface with nearly perfect nesting between straight segments as indicated by dashed magenta lines in Fig. 2(g). They also reported a CDW gap opening of about 20 meV along this Fermi surface. These observations suggest that, besides the p-type higher-order VHS, m-type VHS contributes to the CDW formation as well. The role of the coexistence of distinct types of VHSs in the manifestation of exotic properties in $A$V$_3$Sb$_5$ is an open question.

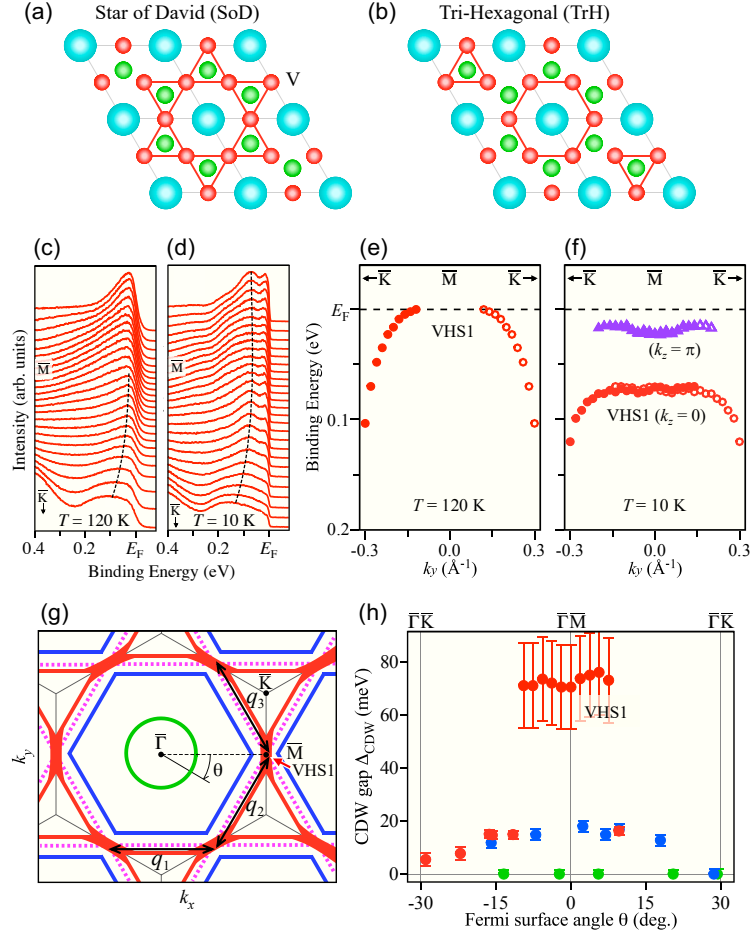

**Fig. 2.** (Color online) (a) and (b) Star of David (SoD) and tri-hexagonal (TrH) lattice distortions, respectively, anticipated for explaining the 2×2 superstructure in the CDW phase of $A$V$_3$Sb$_5$. (c) and (d) ARPES spectra along the MK cut in the normal and CDW state ($T$ = 120 K and 10 K), respectively [16]. Black dashed lines are a guide for the eyes to trace the VHS1 band. (e) and (f) Experimental band dispersion extracted from the peak position in (c) and (d), respectively. (g) Schematic of Fermi surfaces in CsV$_3$Sb$_5$. Red, blue, and magenta lines represent V $d_{x2-y2}$, $d_{yz}$, and $d_{xz}$ orbital characters, respectively, while green shows Sb $p$ orbital character. The existence of extended VHS1 near $E_F$ is expressed by thick red line. Three $q_1$-$q_3$ vectors are indicated by black arrows. Also, the definition of Fermi surface angle θ is displayed. (h) CDW gap size plotted as a function of θ. Red, blue, and green circles show the CDW gap on the V $d_{x2-y2}$, V $d_{yz}$, and Sb $p$ bands, respectively [same color codes as those in (g)]. This figure is reproduced from Ref. [16] (© 2021 American Physical Society).

*3.2 Substitution effect on CDW*

Besides the detailed CDW gap measurements, modulating the band structure, e.g., the energy position of VHSs, is a useful strategy to study the interplay among CDW, superconductivity, and electronic states. Such a study may be carried out by utilizing chemical substitutions. In $A$V$_3$Sb$_5$, replacing V atoms with other transition metal elements modifies CDW and superconducting properties [57-64]. For instance, Ti substitution suppresses CDW, with superconducting transition temperature ($T_c$) reaching a maximum near the CDW phase endpoint [63] [Fig. 3(a)]. Similarly, Nb substitution suppresses CDW and increases $T_c$ [57,64] [Fig. 3(b)]. ARPES measurements of Cs(V,Ti)$_3$Sb$_5$ by Liu *et al.* show a downward energy shift of the chemical potential due to hole doping [63] [Figs. 3(c) and 3(d)], suggesting a scenario that taking VHS1 away from $E_F$ triggers the CDW suppression. In addition, Kato *et al.* [65] reported that V/Nb replacement similarly moves VHS1 away from $E_F$ [Fig. 3(e)], which provides further evidence for a close link between VHS1 and CDW. It is noted that, unlike Ti substitution, Nb substitution does not dope hole carriers. Therefore, to compensate for hole-doping behavior of VHS1, V/Nb substitution results in a downward energy shift of the Sb band at the Γ point [57,65]. The resultant increase of the Sb-derived electron pocket could enhance the density of states at $E_F$ and thus $T_c$.

Like these elemental substitutions in the bulk crystal, surface polarity induces an upward energy shift of VHS1 on the Sb-rich domain of cleaved surfaces in CsV$_3$Sb$_5$, as visualized by using micro-ARPES with a small beam spot [52,53] [see spatial map of domains and band diagram in Figs. 3(g) and 3(h), respectively]. Intriguingly, while CDW-induced band reconstruction is evident on the Cs-rich domain, it is not clearly visible on the Sb-rich domain. Again, this supports an important role of VHS1 for the occurrence of CDW. Investigations into the electronic states near the surface also revealed that heavy electron doping, realized by Cs dosing on cleaved surfaces, suppresses CDW as a result of orbital-selective band shift [66] [see Fig. 3(h), right].

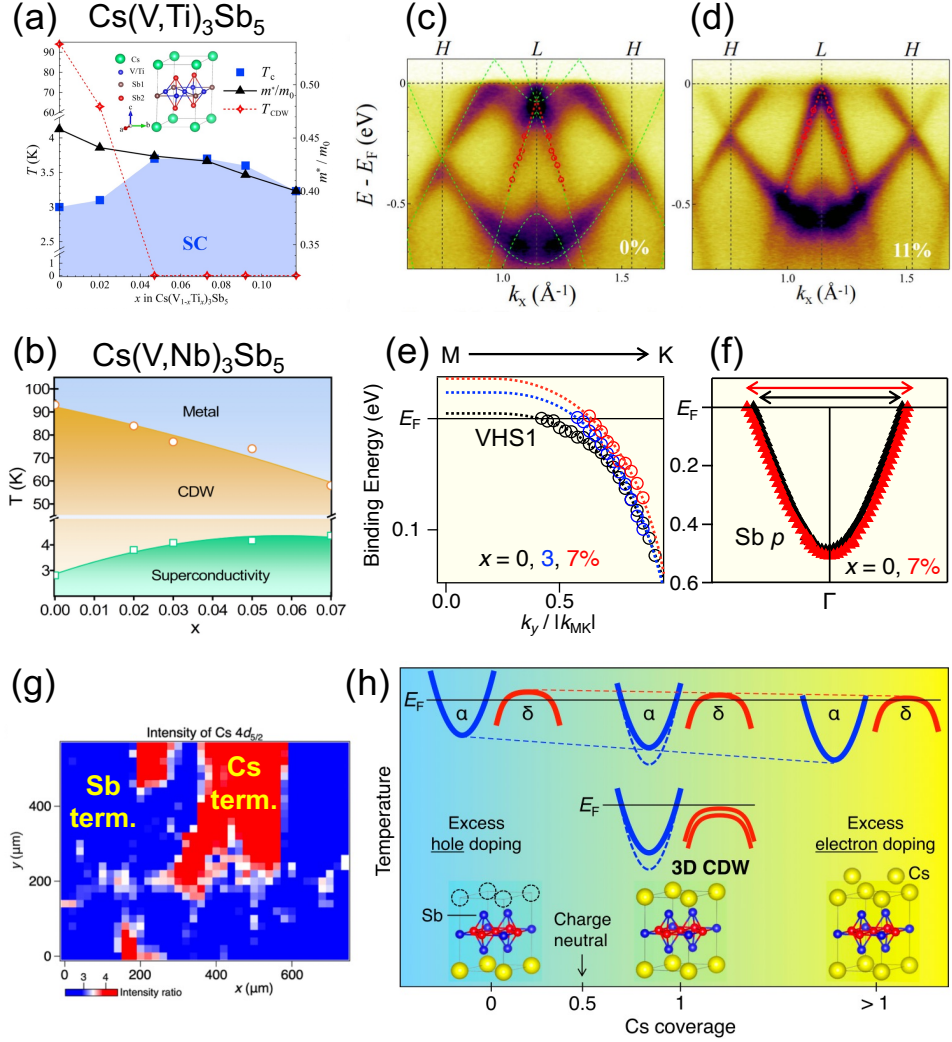

**Fig. 3.** (Color online) (a) and (b) Electronic phase diagram of Cs(V,Ti)$_3$Sb$_5$ [63] and Cs(V,Nb)$_3$Sb$_5$ [57], respectively. (c) and (d) Comparison of VHS dispersions measured in pristine and Ti 11% doped CsV$_3$Sb$_5$ [63]. (e) and (f) Nb-substitution-induced evolution of the band dispersion measured along the MK and ΓK cuts, respectively, in CsV$_3$Sb$_5$ measured at $T$ = 120 K [65]. Black, blue, and red circles represent the band dispersion of pristine, Nb 3%, and Nb 7% samples, respectively. (g) Real-space map of the intensity ratio between surface and bulk Cs 4$d$ core levels, obtained with the step size of 20 μm at $T$ = 8 K [52]. (h) Evolution of band structure in CsV$_3$Sb$_5$ as a function of surface Cs coverage. The inset shows the schematic crystal structure of Sb-terminated (left), Cs-terminated (center), and Cs overdosed (right) CsV$_3$Sb$_5$. Blue and red curves represent the band dispersions around the Γ and M points, respectively. This figure is reproduced from Refs. [52], [57], and [65] (© 2022 American Physical Society), and Ref. [63] (© 2023 The Authors).

# 4. Potential relationship between the electronic states and additional symmetry breakings in the CDW state

*4.1 Role of in-plane 2×2 superstructure*

The CDW state in $A$V$_3$Sb$_5$ not only breaks translational symmetry but also exhibits additional symmetry breakings. These include extra translational symmetry breaking by the development of pair density wave with $4/3a_0$ periodicity in the superconducting state [17], rotational symmetry breaking due to C$_2$-symmetric nematic state [13,20] and 4$a_0$ stripe charge order [12,17,42], and TRS breaking [11,14,15,55,67]. The origins of these symmetry breakings remain largely unexplored. One proposed scenario is that they arise as a secondary effect of the parent CDW transition. This scenario is described below in relation to the CDW-induced band folding.

The SoD or TrH in-plane 2×2 lattice distortion causes Brillouin zone folding [Fig. 4(a)] and band folding from Γ to M, and vice versa, via $q_1$-$q_3$ defined in Fig. 2(g) [Fig. 4(b)]. An interesting consequence of this folding and the associated CDW gap opening is the emergence of small hole pockets [34], which were experimentally found in KV$_3$Sb$_5$ [45] [as indicated by dashed white circles in Fig. 4(c); see also the band dispersion displaying a hole-like character in Fig. 4(d)] and later demonstrated to be a generic feature in $A$V$_3$Sb$_5$ [68] [as specified by green arrow in Fig. 4(e) for CsV$_3$Sb$_5$]. It was proposed that these pockets cause a nesting in the particle-particle channel (but not in the particle-hole channel which leads to CDW or magnetism, since the relevant pockets are all hole-like), resulting in the pair density wave observed below $T_c$ [34]. In fact, the $q_{4/3a0}$ vector connecting these pockets [see green arrow in Fig. 4(f)] is identical to the $q$ vector of the pair density wave [34,68]. This suggests that the electronic reconstruction by the 2×2 CDW not only triggers an additional symmetry breaking at lower temperatures but also establishes an intimate relationship between superconducting and CDW properties. Furthermore, the appearance of the small pockets results in another nesting channel with $q_{4a0}$ denoted by black arrow in Fig. 4(f), suggesting a connection with the 4$a_0$-period stripe charge order observed in STM [12,17,42]. However, as the stripe order has not been detected in bulk-sensitive probes, it may originate from surface reconstruction [56,69]. Therefore, further study is required to elucidate its origin.

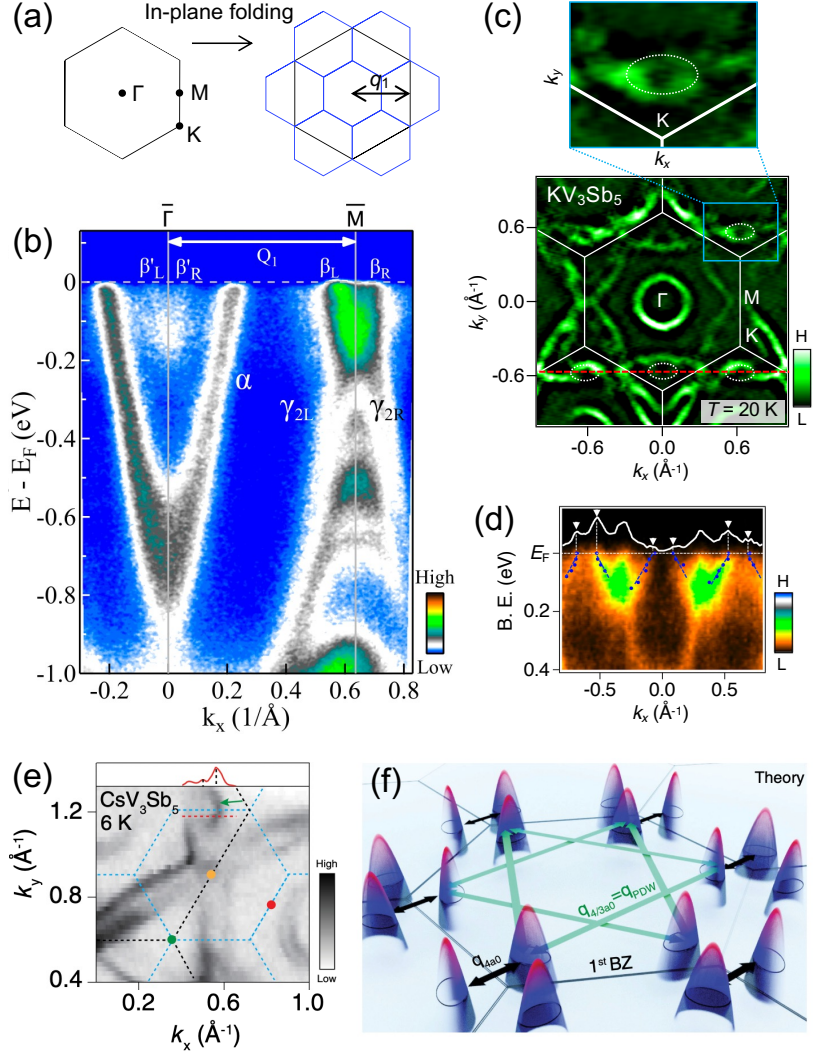

**Fig. 4.** (Color online) (a) Schematic of Brillouin zone folding due to the in-plane 2×2 superlattice potential. (b) Plot of ARPES intensity along the $\overline{\Gamma M}$ line measured in the CDW phase ($T$ = 20 K) in CsV$_3$Sb$_5$ [49]. The β bands which are originally located at the $\overline{M}$ point are folded onto the $\overline{\Gamma}$ point. (c) Plot of second-derivative ARPES intensity at $E_F$ measured in the CDW phase ($T$ = 20 K) in KV$_3$Sb$_5$ [45]. Small hole pockets induced by the CDW-induced reconstruction are highlighted by white dashed ellipses. (d) Band dispersion measured along red dashed line in (c). Blue dashed lines are a guide for the eyes to trace the small pocket. The inset shows the momentum distribution curve at $E_F$, in which peak position marked by white triangle corresponds to the Fermi wave vector of the small pocket. (e) Reconstructed Fermi surface in CsV$_3$Sb$_5$. The small pocket is marked by green arrow [68]. (f) Schematic illustration of new scattering channels originating from the emergence of small hole pockets. This figure is reproduced from Refs. [45] and [49] (© 2022 The Authors), and Ref. [68] (© 2023 American Physical Society).

*4.2 Role of 3D CDW*

In $A$V$_3$Sb$_5$, in addition to the in-plane unit-cell expansion, CDW typically generates a unit-cell doubling along the $c$ axis [42,43,69], causing Brillouin zone folding in the $k_z$ direction [Fig. 5(a)]. This 3D aspect may result from an alternative stacking of SoD and TrH structures [32,50] [Fig. 5(b)] or a π-phase shift of the SoD or TrH structure between adjacent kagome planes [Figs. 5(c) and 5(d)]. Naively, these stacking models lead to band doubling (splitting) due to the band folding from $k_z$ = 0 to π, and vice versa, for a band showing finite $k_z$ dispersion, like VHS1 [see Fig. 5(e)]. Calculations indeed reproduce the splitting of VHS1 and other bands, but the degree of splitting varies depending on the distortion type and the presence/absence of a π-phase shift [50] [compare Figs. 5(f)-5(h)]. Experimentally, the splitting was discovered for the VHS1 band, the Dirac cone band, and the band bottomed at the M point in the CDW phase of CsV$_3$Sb$_5$ [50,51] [Figs. 5(i) and 5(j)], confirming 3D-CDW-induced electronic reconstruction. The observed band splitting is best compatible with the alternating SoD and TrH model in CsV$_3$Sb$_5$, while staggered 3D CDW appears to be realized in KV$_3$Sb$_5$ [50] (in RbV$_3$Sb$_5$, the possibilities of both alternating and staggered phases were reported [50]). It is noted that, while 2×2×4 phase is also suggested in CsV$_3$Sb$_5$ [44,70], the expected quadrupling of the band structure has not been clearly observed, possibly due to the fragility of this phase.

An essential difference between the alternating SoD and TrH [Fig. 5(b)] and staggered SoD or TrH [Figs. 5(c) and 5(d)] is that the latter breaks the original C$_6$ symmetry of the kagome lattice [32,71], but the former does not. This is evident from a C$_2$ symmetric arrangement of atoms in a projected crystal structure in Fig. 5(k). This symmetry lowering possibly contributes to the development of nematicity in the CDW phase. Notably, a recent ARPES study by Jiang *et al*. [72] reported the C$_2$ symmetric behavior of the Fermi surface in KV$_3$Sb$_5$, as recognized from one-dimensional intensity distributions elongated exclusively along two of the three ΓM lines (ΓM$_2$ and ΓM$_3$ lines) in Figs. 5(l) and 5(m). This result again suggests the notion that CDW-induced electronic and lattice reconstructions trigger additional symmetry breakings inside the CDW phase.

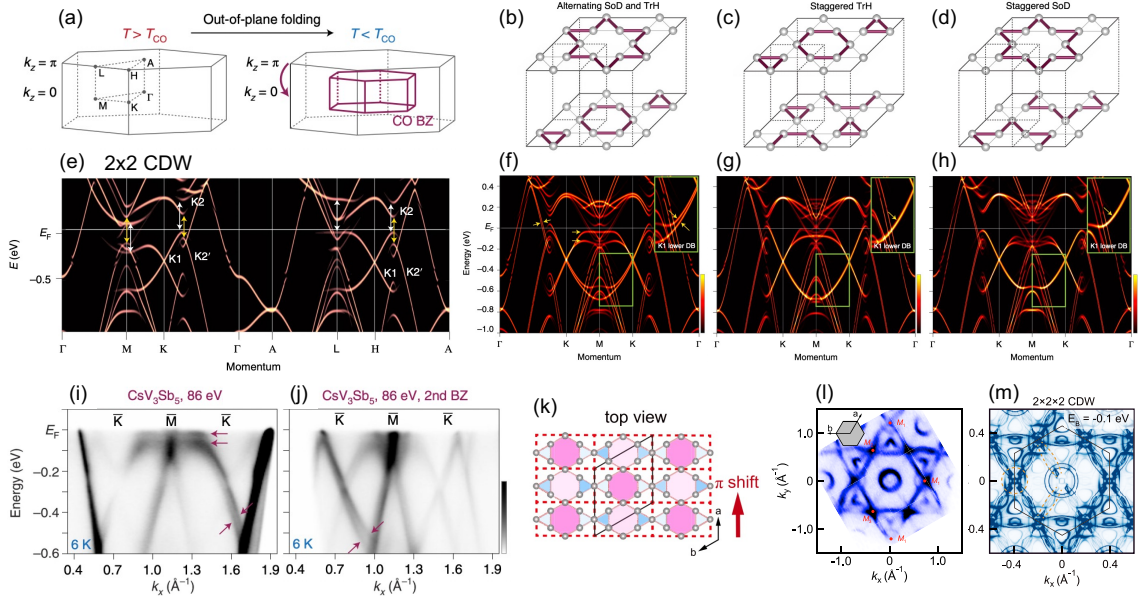

**Fig. 5.** (Color online) (a) Schematic of Brillouin zone folding caused by 2×2×2 CDWs. (b)-(d) Schematic crystal structure in the 2×2×2 CDW phase consisting of (b) alternate stacks of SoD- and TrH-distorted V kagome layers, (c) TrH-distorted layers with a π-phase shift, and (d) SoD-distorted layers with a π-phase shift. (e) Unfolded band structure calculated by DFT assuming 2×2 CDW. White arrows at the M and L points highlight a CDW gap opening on the VHS1 band. (f)-(h) DFT calculations for the 2×2×2 CDW phases in (b)-(d). Yellow arrows in (f) signify the band splitting. Differences in the presence/absence of the band splitting among the three models are identified in the regions enclosed by green rectangles. (i) and (j) ARPES intensity plots measured along the MK direction in different Brillouin zones (1st vs. 2nd zones). Because of the difference in matrix element effects, the intensity of different parts of the band structure is enhanced between the two results. The band splitting is marked by arrows. Figs. (a)-(j) are reproduced from Ref. [50] (© 2023 The Authors). (k) Top view of the crystal structure in the staggered TrH CDW [72]. (l) ARPES intensity plot at a binding energy of 0.2 eV measured in the CDW state ($T$ = 14 K) in $KV_3Sb_5$. (m) Equi-energy contour plot at a binding energy of 0.1 eV obtained by DFT calculations assuming the staggered TrH CDW. Yellow dashed ellipses highlight the two-fold symmetric electronic states. Figs. (k)-(m) are reproduced from Ref. [72] (© 2023 American Chemical Society).

## 5. Superconductivity

In $A$V$_3$Sb$_5$ compounds, the superconductivity that intertwines with CDW emerges at $T_c$ of around 1 K for KV$_3$Sb$_5$ and RbV$_3$Sb$_5$, and around 3 K for CsV$_3$Sb$_5$. As theoretically predicted [30,33,48,73,74], the superconductivity emergent in the kagome lattice could exhibit an unconventional chirality. Therefore, studying the nature of the superconductivity of $A$V$_3$Sb$_5$ is important to identify the potential chiral superconductivity. In this context, ARPES is useful to study the many-body interactions, such as electron-phonon coupling (EPC), and directly determine the superconducting (SC) gap in momentum space. In this section, we present the ARPES progress on these two regards and discuss possible pairing mechanisms.

*5.1 Electron-phonon coupling and its relationship with superconductivity*

Studying the EPC is an essential step for a new superconductor to decode the origin of superconductivity and its interplay with other electronic orders. Intuitively, in $A$V$_3$Sb$_5$, multiple VHSs from V 3$d$-electrons near $E_F$ revealed by the ARPES studies of normal states [38,39] [Figs. 1(f) and 1(h)], as well as that the DFT calculated EPC strength [26], ~ 0.25, fails to support the $T_c$, together highlighting the electronic driven instabilities [27,48]. However, experimentally the EPC is proven to be not negligible and can support the superconducting transitions [19,49,75-77].

The ARPES study by Luo *et al.* [49] of a cousin compound KV$_3$Sb$_5$ first reported a clear kink below $E_F$ in the electronic band structure near the VHS, suggesting a moderate EPC. The later ARPES study of CsV$_3$Sb$_5$ by Zhong *et al.* [19] further gives the orbital- and momentum-dependence of the EPC strength and tests the relationship between EPC and $T_c$. As shown in Figs. 6(a)-6(c), one dominated kink was observed on the α band [occupied by Sb-5$p$ electrons which form the circular Fermi surface shown by red color in Fig.6(a)], while two prominent kinks were observed on the β band [occupied by V-3$d_{yz}$ electrons which form the hexagonal Fermi surface shown by blue color in Fig. 6(a)]. Despite of this, the EPC strength estimated from these kinks is similar for both Sb 5$p$ and V 3$d$ electronic bands and located in an intermediate range of 0.45 ~ 0.6. Moreover, the momentum-dependent study shows a nearly isotropic EPC, as shown in Fig. 6(f). Importantly, by extracting Eliashberg functions shown in Fig. 6(e), such an isotropic EPC is proven to support the SC transition at a temperature on the same magnitude of the experimental $T_c$, highlighting the important role of EPC on the superconductivity of CsV$_3$Sb$_5$. This is further emphasized by the experimental facts [18,19] that the EPC on V 3$d$-band is enhanced by about 50% in the isovalent-substituted Cs(V$_{0.93}$Nb$_{0.07}$)$_3$Sb$_5$ with an elevated $T_c \simeq 4.4$ K, as well as in Cs(V$_{0.86}$Ta$_{0.14}$)$_3$Sb$_5$ with an elevated $T_c \simeq 5.2$ K [Figs.

6(g)–6(i)].

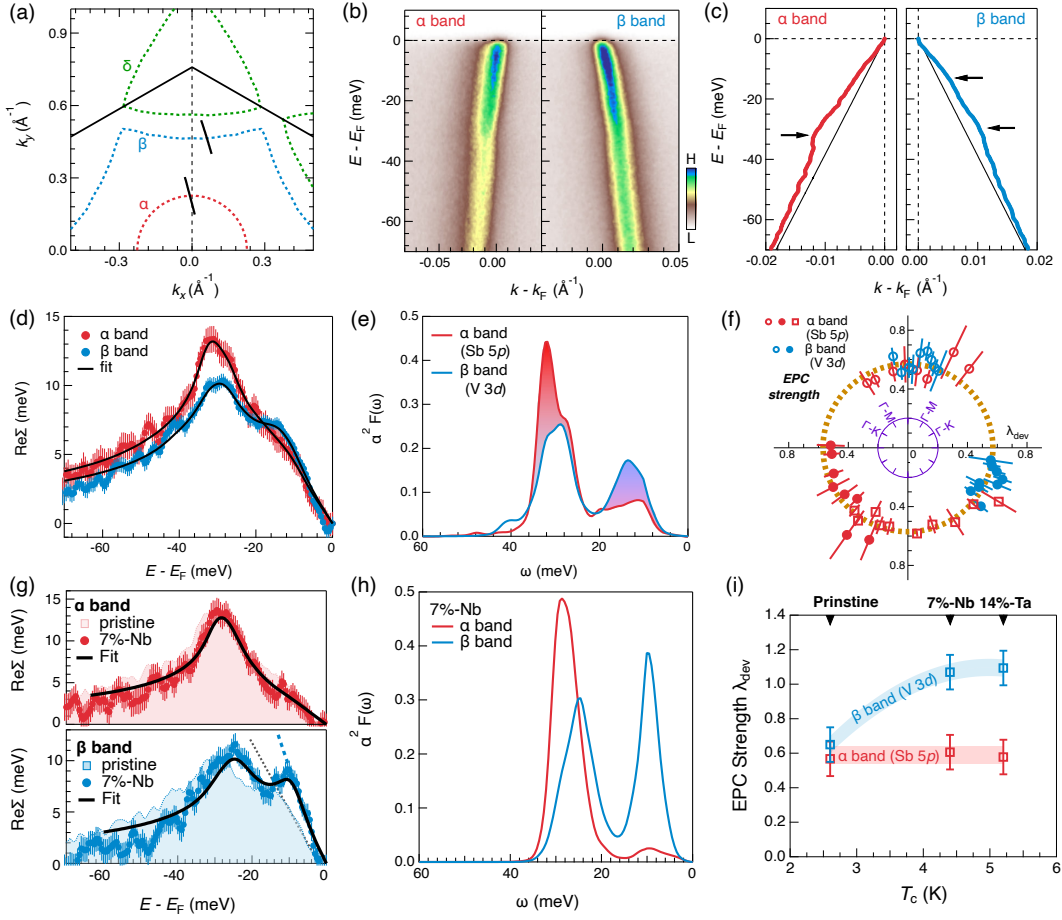

**Fig. 6.** (Color online) (a) Contours of the FS of $CsV_3Sb_5$. (b) ARPES intensity plots of the α and β bands. (c) Extracted band dispersions of the α and β bands. The arrows show the energy position of the kinks, and the black lines are the bare bands. (d) Real part of self-energy ReΣ and the superimposed black lines are ReΣ reproduced by maximum entropy method. (e) Eliashberg function deduced from ReΣ. (f) EPC strength defined by the slope of ReΣ at $E_F$ plotted with FS angle. The data presented in (b)-(f) were taken from pristine $CsV_3Sb_5$. (g) Real part of self-energy of 7%-Nb doped sample, $Cs(V_{0.93}Nb_{0.07})_3Sb_5$, and the ReΣ of pristine $CsV_3Sb_5$ is plotted as colored shadows for a direct comparison. (h) Eliashberg function of 7%-Nb doped sample. (i) Summary of the EPC strength estimated from the derivative of ReΣ at $E_F$ for pristine $CsV_3Sb_5$, 7%-Nb and 14%-Ta doped $CsV_3Sb_5$. Figure is reproduced from Ref. [19] (© 2023 The Authors).

Besides the APRES observation of the electronic kinks, the existence of a strong EPC are individually confirmed by the observations of phonon hardening effect in neutron scattering measurements [75], strong phonon anomalies in optical spectroscopy [76] and Raman scattering measurements [77].

*5.2 Superconducting gap symmetry*

SC gap symmetry is fundamentally crucial to illuminate the pairing mechanism. However, it remains elusive owing to the existence of several conflicting experimental results. The observations of the certain V-shaped gaps, as well as residual Fermi-level states measured by scanning tunnelling spectroscopy [17,78] and a finite residual thermal conductivity towards zero temperature [79] in $CsV_3Sb_5$ supports a nodal SC gap scenario. By contrast, the observations of the Hebel–Slichter coherence peak in the spin-lattice relaxation rate from $^{121/123}$Sb nuclear quadrupole resonance measurements [80] and the exponentially temperature-dependent magnetic penetration depth [81,82], are more consistent with a nodeless SC gap symmetry. Furthermore, the muon spin relaxation (μSR) measurements [83] on $RbV_3Sb_5$ and $KV_3Sb_5$ superconductors reported a transition from the nodal to nodeless superconductivity by suppressing the charge order with applying pressure. Theoretically, both nodal and nodeless pairing symmetries were proposed depending on the situations of electron interactions [30,33,84].

In the past decades, ARPES has been proved to be a powerful tool to directly measure the SC gap in the momentum space [85,86]. Despite an ARPES study for $CsV_3Sb_5$ observed the same band gap for α and β bands [87], the momentum dependence of the SC gap is lacking due to the relatively low $T_c$ and resolution challenges on determining a tiny SC gap. Zhong *et al.* [18] performed ultra-high and low-temperature laser-based ARPES measurements on two chemically substituted $CsV_3Sb_5$ samples, $Cs(V_{0.93}Nb_{0.07})_3Sb_5$ and $Cs(V_{0.86}Ta_{0.14})_3Sb_5$ that raises $T_c$ to 4.4 K and 5.2 K, respectively. The ARPES spectra for three Fermi surfaces of $Cs(V_{0.86}Ta_{0.14})_3Sb_5$ are shown in Fig. 7(a) and the SC gap magnitudes extracted from these ARPES spectra are summarized in Figs. 7(b)-7(c) [note that the δ Fermi surface has V $d_{x2-y2}$ character producing VHS1, indicated by red in Fig. 2(g)]. Clearly, the SC gap in $Cs(V_{0.86}Ta_{0.14})_3Sb_5$ is nodeless, nearly isotropic and orbital-independent. Interesting point is that $Cs(V_{0.86}Ta_{0.14})_3Sb_5$ exhibits a $T_c$ of 5.2 K, but no clear CDW transition, shown as Fig. 7(d). To examine the influence of CDW, the SC gap in $Cs(V_{0.93}Nb_{0.07})_3Sb_5$ which exhibits a similar $T_c$ (~ 4.4 K) but a partially suppressed CDW transition ($T_{CDW}$ ~ 58 K) was further studied. Surprisingly, the high-resolution ARPES results shows the robust nodeless, nearly isotropic and orbital-independent SC gap is remained in charge-ordered $Cs(V_{0.93}Nb_{0.07})_3Sb_5$ compound [18] [Figs. 7(e)-7(g)].

These results provide indispensable information on the electron pairing symmetry of kagome superconductors. The robust isotropic SC gaps in the presence or absence of the CDW and the enough strong EPC tend to a conventional *s*-wave pairing scenario. Precisely, other nodeless pairing states [30,33], such as chiral *p*+i*p* or *s*+i*s* waves, are also consistent with the ARPES observations [18,87]. Especially, the presence of TRS inside the SC state was observed by μSR measurements of pressurized CsV$_3$Sb$_5$ in which CDW order is eliminated [88].

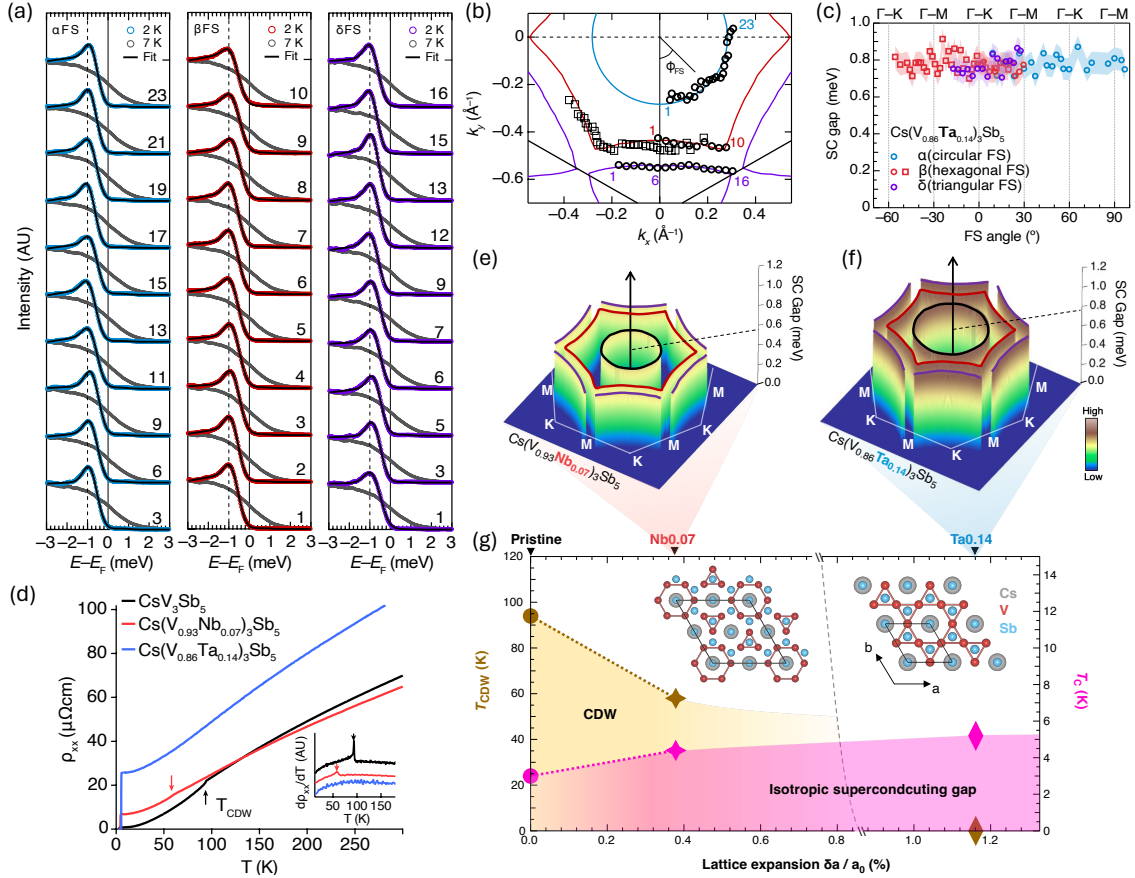

**Fig. 7.** (Color online) (a) Represented EDCs of 14%-Ta doped sample, Cs(V$_{0.86}$Ta$_{0.14}$)Sb$_5$, at $k_F$ of α, β and δ FSs. (b) FS contours of 14%-Ta doped sample and the markers show the $k_F$ positions where the SC gap was measured. (c) Extracted SC gap magnitude plotted as function of FS angle defined as in (b). (d) Resistivity curves of pristine CsV$_3$Sb$_5$, 7%-Nb and 14%-Ta doped CsV$_3$Sb$_5$. The inset is the derivative for clarifying the CDW transition. (e) SC gap distribution in the momentum space of 7%-Nb doped CsV$_3$Sb$_5$ determined by high-resolution ARPES measurements. (f) same as (e) but for 14%-Ta doped CsV$_3$Sb$_5$. (g) Schematic phase diagram in which $T$ is plotted with lattice expansion $\delta a = (a-a_0)/a_0$ due to the chemical substitutions. $a_0$ is the in-plane lattice constant of pristine CsV$_3$Sb$_5$. The inset shows the lattice structures of the CDW (left) and undistorted (right) phases, representing the states above $T_c$ for two regions in the phase diagram. Figure is reproduced from Ref. [18] (© 2023 The Authors).

*5.3 Possible superconducting mechanism*

In $A$V$_3$Sb$_5$ family materials, the nonlocal electronic correlation effect through the sublattice interference harbored by multiple VHSs associated with V 3$d$ orbitals [38,39] is important for the appearance of intriguing phenomena, as mentioned in Section 3. For superconductivity, this correlation will give rise to a nodal/nodeless pairing with strong anisotropy or a chiral $d$/$p$-wave pairing, depending on the extent of the correlation [30,33]. In particular, the chiral $d$/$p$-wave pairing is nodeless and breaks TRS, consistent with ARPES [18] and μSR observations [88]. However, in the scenario of the pure electronic interaction [30,33], the high density of states near $E_F$ at M point, due to the VHSs associated with V 3$d$ orbitals, is expected to result in a strongly anisotropic SC gap on V 3$d$ orbitals and different gaps between V 3$d$ and Sb 5$p$ orbitals [34,89]. These predictions are partially inconsistent with orbital-independent and isotropic SC gap revealed by ARPES. On the other hand, EPC is experimentally proven to be non-negligible and positively correlated with $T_c$ [19]. While EPC alone typically generates a phase-uniform $s$-wave pairing, it does not account for the TRS breaking in SC state. Moreover, the distinct EPC strength inferred from the electronic kink dispersion suggests that EPC-driven pairing could also produce different SC gaps on V 3$d$ and Sb 5$p$ orbitals [Fig. 6(i)]. Thus, while both electronic interaction and EPC play crucial roles in promoting superconductivity in these kagome superconductors, neither of them alone can fully explain the formation of the superconductivity. Instead, they may reinforce each other in the frustrated kagome lattice [90], collaboratively generating a TRS broken pairing with an isotropic gap, which warrants further theoretical investigations. We notice that the element substitutions in V site could induce a change in the SC gap symmetry [33]. Further considering the similar EPC strength on V and Sb atomic orbitals in pristine CsV$_3$Sb$_5$ [Fig. 6(i)], in the future, a direct ARPES investigation on pristine CsV$_3$Sb$_5$ will be more helpful to further pinpointing the pairing symmetry, then deepening our understanding on the pairing mechanism of $A$V$_3$Sb$_5$ kagome superconductors.

**6. Conclusion**

In conclusion, our review of recent advancements in photoemission studies of the kagome superconductor $A$V$_3$Sb$_5$ has provided valuable insights into the unconventional CDW order and electron pairing symmetry. Although the origins of CDW order remain contentious, the intimate connection between CDW order and VHSs is strongly supported by the photoemission observations, which reveal that the maximum CDW gap occurs at a VHS and its energy shifts accordingly with the suppression of CDW. On superconductivity front, the observations of a robust nodeless SC gap and moderate EPC suggest a tendency towards conventional superconductivity. However, the possibility of an exotic chiral superconductivity cannot be entirely ruled out, particularly considering the potential TRS breaking in SC state as suggested by μSR studies. These findings collectively underscore the significant influence of electron correlations in driving the observed electronic orders in $A$V$_3$Sb$_5$. Nevertheless, the intriguing complexities of this system warrant further ARPES investigations, including electronic structure of lower symmetry in the CDW, signature of the TRS breaking, the SC gap of pristine $A$V$_3$Sb$_5$, and the interplay between the CDW and superconductivity, holding a promise of a deeper understanding of kagome-related physics.

**Acknowledgments**

The authors thank Dr. Takemi Kato, Prof. Takafumi Sato, Prof. Kozo Okazaki, Prof. Zhiwei Wang, and Prof. Yugui Yao for their invaluable contributions and fruitful discussions during collaborative works. Y. Zhong thanks the Japan Society for the Promotion of Science (JSPS) for the funding through the *JSPS International Research Fellowship* program (KAKENHI No. 24KF0021). K. Nakayama is supported by JSPS (KAKENHI No. 23H01115).

**References**

[1] B. Keimer, S. A. Kivelson, M. R. Norman, S. Uchida, and J. Zaanen, From quantum matter to high-temperature superconductivity in copper oxides, *Nature* **518**, 179 (2015).

[2] R. M. Fernandes, A. I. Coldea, H. Ding, I. R. Fisher, P. Hirschfeld, and G. Kotliar, Iron pnictides and chalcogenides: a new paradigm for superconductivity, *Nature* **601**, 35 (2022).

[3] X. L. Qi and S. C. Zhang, Topological insulators and superconductors, *Reviews of Modern Physics* **83**, 1057 (2011).

[4] B. Q. Lv, T. Qian, and H. Ding, Experimental perspective on three-dimensional topological semimetals, *Reviews of Modern Physics* **93**, 025002 (2021).

[5] I. Syôzi, Statistics of kagomé lattice, *Progress of Theoretical Physics* **6**, 306 (1951).

[6] J.-X. Yin, B. Lian, and M. Z. Hasan, Topological kagome magnets and superconductors, *Nature* **612**, 647 (2022).

[7] T. Neupert, M. M. Denner, J.-X. Yin, R. Thomale, and M. Z. Hasan, Charge order and superconductivity in kagome materials, *Nature Physics* **18**, 137 (2021).

[8] K. Jiang, T. Wu, J. X. Yin, Z. Y. Wang, M. Z. Hasan, S. D. Wilson, X. H. Chen, and J. P. Hu, Kagome superconductors $A$V$_3$Sb$_5$ ($A$ = K, Rb, Cs), *Natl Sci Rev* **10**, nwac199 (2023).

[9] B. R. Ortiz, L. C. Gomes, J. R. Morey, M. Winiarski, M. Bordelon, J. S. Mangum, I. W. H. Oswald, J. A. Rodriguez-Rivera, J. R. Neilson, S. D. Wilson *et al.*, New kagome prototype materials: discovery of KV$_3$Sb$_5$, RbV$_3$Sb$_5$, and CsV$_3$Sb$_5$, *Phys. Rev. Mat.* **3**, 094407 (2019).

[10] B. R. Ortiz, S. M. L. Teicher, Y. Hu, J. L. Zuo, P. M. Sarte, E. C. Schueller, A. M. M. Abeykoon, M. J. Krogstad, S. Rosenkranz, R. Osborn *et al.*, CsV$_3$Sb$_5$: A $Z_2$ Topological Kagome Metal with a Superconducting Ground State, *Phys. Rev. Lett.* **125**, 247002 (2020).

[11] Y. X. Jiang, J. X. Yin, M. M. Denner, N. Shumiya, B. R. Ortiz, G. Xu, Z. Guguchia, J. He, M. S. Hossain, X. Liu *et al.*, Unconventional chiral charge order in kagome superconductor KV$_3$Sb$_5$, *Nat Mater* **20**, 1353 (2021).

[12] H. Zhao, H. Li, B. R. Ortiz, S. M. L. Teicher, T. Park, M. Ye, Z. Wang, L. Balents, S. D. Wilson, and I. Zeljkovic, Cascade of correlated electron states in the kagome superconductor CsV$_3$Sb$_5$, *Nature* **599**, 216 (2021).

[13] L. P. Nie, K. Sun, W. R. Ma, D. W. Song, L. X. Zheng, Z. W. Liang, P. Wu, F. H. Yu, J. Li, M. Shan *et al.*, Charge-density-wave-driven electronic nematicity in a kagome superconductor, *Nature* **604**, 59 (2022).

[14] C. Mielke, III, D. Das, J. X. Yin, H. Liu, R. Gupta, Y. X. Jiang, M. Medarde, X. Wu, H. C. Lei, J. Chang *et al.*, Time-reversal symmetry-breaking charge order in a kagome

superconductor, *Nature* **602**, 245 (2022).

[15] Y. Xu, Z. Ni, Y. Liu, B. R. Ortiz, Q. Deng, S. D. Wilson, B. Yan, L. Balents, and L. Wu, Three-state nematicity and magneto-optical Kerr effect in the charge density waves in kagome superconductors, *Nature Physics* **18**, 1470 (2022).

[16] K. Nakayama, Y. Li, T. Kato, M. Liu, Z. Wang, T. Takahashi, Y. Yao, and T. Sato, Multiple energy scales and anisotropic energy gap in the charge-density-wave phase of the kagome superconductor CsV$_3$Sb$_5$, *Physical Review B* **104**, L161112 (2021).

[17] H. Chen, H. Yang, B. Hu, Z. Zhao, J. Yuan, Y. Xing, G. Qian, Z. Huang, G. Li, Y. Ye *et al.*, Roton pair density wave in a strong-coupling kagome superconductor, *Nature* **599**, 222 (2021).

[18] Y. G. Zhong, J. J. Liu, X. X. Wu, Z. Guguchia, J. X. Yin, A. Mine, Y. K. Li, S. Najafzadeh, D. Das, C. Mielke *et al.*, Nodeless electron pairing in CsV$_3$Sb$_5$-derived kagome superconductors, *Nature* **617**, 488 (2023).

[19] Y. G. Zhong, S. Z. Li, H. X. Liu, Y. Y. Dong, K. Aido, Y. Arai, H. X. Li, W. L. Zhang, Y. G. Shi, Z. Q. Wang *et al.*, Testing electron-phonon coupling for the superconductivity in kagome metal Cs$_3$VSb$_5$, *Nat. Commun.* **14**, 1945 (2023).

[20] Y. Xiang, Q. Li, Y. Li, W. Xie, H. Yang, Z. Wang, Y. Yao, and H.-H. Wen, Twofold symmetry of c-axis resistivity in topological kagome superconductor CsV$_3$Sb$_5$ with in-plane rotating magnetic field, *Nat. Commun.* **12**, 6727 (2021).

[21] H. Li, H. Zhao, B. R. Ortiz, T. Park, M. Ye, L. Balents, Z. Wang, S. D. Wilson, and I. Zeljkovic, Rotation symmetry breaking in the normal state of a kagome superconductor KV$_3$Sb$_5$, *Nature Physics* **18**, 265 (2022).

[22] S. Y. Yang, Y. Wang, B. R. Ortiz, D. Liu, J. Gayles, E. Derunova, R. Gonzalez-Hernandez, L. Smejkal, Y. Chen, S. S. P. Parkin *et al.*, Giant, unconventional anomalous Hall effect in the metallic frustrated magnet candidate, KV$_3$Sb$_5$, *Science Advances* **6**, eabb6003 (2020).

[23] F. H. Yu, T. Wu, Z. Y. Wang, B. Lei, W. Z. Zhuo, J. J. Ying, and X. H. Chen, Concurrence of anomalous Hall effect and charge density wave in a superconducting topological kagome metal, *Physical Review B* **104**, L041103 (2021).

[24] K. Y. Chen, N. N. Wang, Q. W. Yin, Y. H. Gu, K. Jiang, Z. J. Tu, C. S. Gong, Y. Uwatoko, J. P. Sun, H. C. Lei *et al.*, Double superconducting dome and triple enhancement of $T_c$ in the kagome superconductor CsV$_3$Sb$_5$ under high pressure, *Phys. Rev. Lett.* **126**, 247001 (2021).

[25] F. Yu, D. Ma, W. Zhuo, S. Liu, X. Wen, B. Lei, J. Ying, and X. Chen, Unusual competition of superconductivity and charge-density-wave state in a compressed topological kagome metal, *Nat. Commun.* **12**, 3645 (2021).

[26] H. Tan, Y. Liu, Z. Wang, and B. Yan, Charge Density Waves and Electronic Properties of Superconducting Kagome Metals, *Phys. Rev. Lett.* **127**, 046401 (2021).

[27] T. Park, M. Ye, and L. Balents, Electronic instabilities of kagome metals: saddle points and Landau theory, *Physical Review B* **104**, 035142 (2021).

[28] Y. P. Lin and R. M. Nandkishore, Complex charge density waves at Van Hove singularity on hexagonal lattices: Haldane-model phase diagram and potential realization in the kagome metals $A$V$_3$Sb$_5$ ($A$ = K, Rb, Cs), *Physical Review B* **104** (2021).

[29] M. M. Denner, R. Thomale, and T. Neupert, Analysis of Charge Order in the Kagome Metal $A$V$_3$Sb$_5$ ($A$=K, Rb, Cs), *Phys. Rev. Lett.* **127**, 217601 (2021).

[30] X. Wu, T. Schwemmer, T. Müller, A. Consiglio, G. Sangiovanni, D. Di Sante, Y. Iqbal, W. Hanke, A. P. Schnyder, M. M. Denner *et al.*, Nature of Unconventional Pairing in the Kagome Superconductors $A$V$_3$Sb$_5$ ($A$=K, Rb, Cs), *Phys. Rev. Lett.* **127**, 177001 (2021).

[31] X. Feng, K. Jiang, Z. Wang, and J. Hu, Chiral flux phase in the Kagome superconductor $A$V$_3$Sb$_5$, *Science bulletin* **66**, 1384 (2021).

[32] M. H. Christensen, T. Birol, B. M. Andersen, and R. M. Fernandes, Theory of the charge density wave in CsV$_3$Sb$_5$ kagome metals, *Physical Review B* **104**, 214513 (2021).

[33] R. Tazai, Y. Yamakawa, S. Onari, and H. Kontani, Mechanism of exotic density-wave and beyond-Migdal unconventional superconductivity in kagome metal $A$V$_3$Sb$_5$ ($A$= K, Rb, Cs), *Science Advances* **8**, eabl4108 (2022).

[34] S. Zhou and Z. Q. Wang, Chern Fermi pocket, topological pair density wave, and charge-4e and charge-6e superconductivity in kagome superconductors, *Nat. Commun.* **13**, 7288 (2022).

[35] R. Tazai, Y. Yamakawa, and H. Kontani, Charge-loop current order and nematicity mediated by bond order fluctuations in kagome metals, *Nat. Commun.* **14**, 7845 (2023).

[36] J. W. Dong, Z. Q. Wang, and S. Zhou, Loop-current charge density wave driven by long-range Coulomb repulsion on the kagome lattice, *Physical Review B* **107**, 045127 (2023).

[37] M. L. Kiesel and R. Thomale, Sublattice interference in the kagome Hubbard model, *Physical Review B* **86**, 121105 (2012).

[38] M. G. Kang, S. A. Fang, J. K. Kim, B. R. Ortiz, S. H. Ryu, J. M. Kim, J. Yoo, G. Sangiovanni, D. Di Sante, B. G. Park *et al.*, Twofold van Hove singularity and origin of charge order in topological kagome superconductor CsV$_3$Sb$_5$, *Nature Physics* **18**, 301 (2022).

[39] Y. Hu, X. Wu, B. R. Ortiz, S. Ju, X. Han, J. Ma, N. C. Plumb, M. Radovic, R. Thomale, S. D. Wilson *et al.*, Rich nature of Van Hove singularities in Kagome

superconductor CsV$_3$Sb$_5$, *Nat. Commun.* **13**, 2220 (2022).

[40] N. F. Q. Yuan, H. Isobe, and L. Fu, Magic of high-order van Hove singularity, *Nat. Commun.* **10**, 5769 (2019).

[41] L. Classen, A. V. Chubukov, C. Honerkamp, and M. M. Scherer, Competing orders at higher-order Van Hove points, *Physical Review B* **102**, 125141 (2020).

[42] Z. Liang, X. Hou, F. Zhang, W. Ma, P. Wu, Z. Zhang, F. Yu, J. J. Ying, K. Jiang, L. Shan *et al.*, Three-Dimensional Charge Density Wave and Surface-Dependent Vortex-Core States in a Kagome Superconductor CsV$_3$Sb$_5$, *Physical Review X* **11**, 031026 (2021).

[43] H. Li, T. T. Zhang, T. Yilmaz, Y. Y. Pai, C. E. Marvinney, A. Said, Q. W. Yin, C. S. Gong, Z. J. Tu, E. Vescovo *et al.*, Observation of Unconventional Charge Density Wave without Acoustic Phonon Anomaly in Kagome Superconductors $A$V$_3$Sb$_5$ ($A$= Rb, Cs), *Physical Review X* **11**, 031050 (2021).

[44] B. R. Ortiz, S. M. L. Teicher, L. Kautzsch, P. M. Sarte, N. Ratcliff, J. Harter, J. P. C. Ruff, R. Seshadri, and S. D. Wilson, Fermi Surface Mapping and the Nature of Charge-Density-Wave Order in the Kagome Superconductor CsV$_3$Sb$_5$, *Physical Review X* **11**, 041030 (2021).

[45] T. Kato, Y. K. Li, T. Kawakami, M. Liu, K. Nakayama, Z. W. Wang, A. Moriya, K. Tanaka, T. Takahashi, Y. G. Yao *et al.*, Three-dimensional energy gap and origin of charge-density wave in kagome superconductor KV$_3$Sb$_5$, *Commun Mater* **3**, 30 (2022).

[46] Z. Liu, N. Zhao, Q. Yin, C. Gong, Z. Tu, M. Li, W. Song, Z. Liu, D. Shen, Y. Huang *et al.*, Charge-Density-Wave-Induced Bands Renormalization and Energy Gaps in a Kagome Superconductor RbV$_3$Sb$_5$, *Physical Review X* **11**, 041010 (2021).

[47] W.-S. Wang, Z.-Z. Li, Y.-Y. Xiang, and Q.-H. Wang, Competing electronic orders on kagome lattices at van Hove filling, *Physical Review B* **87**, 115135 (2013).

[48] M. L. Kiesel, C. Platt, and R. Thomale, Unconventional fermi surface instabilities in the kagome Hubbard model, *Phys. Rev. Lett.* **110**, 126405, 126405 (2013).

[49] H. Luo, Q. Gao, H. Liu, Y. Gu, D. Wu, C. Yi, J. Jia, S. Wu, X. Luo, Y. Xu *et al.*, Electronic nature of charge density wave and electron-phonon coupling in kagome superconductor KV$_3$Sb$_5$, *Nat. Commun.* **13**, 273 (2022).

[50] M. G. Kang, S. Fang, J. Yoo, B. R. Ortiz, Y. M. Oey, J. Choi, S. H. Ryu, J. Kim, C. Jozwiak, A. Bostwick *et al.*, Charge order landscape and competition with superconductivity in kagome metals, *Nat. Mater.* **22**, 186 (2023).

[51] Y. Hu, X. X. Wu, B. R. Ortiz, X. L. Han, N. C. Plumb, S. D. Wilson, A. P. Schnyder, and M. Shi, Coexistence of trihexagonal and star-of-David pattern in the charge density wave of the kagome superconductor $A$V$_3$Sb$_5$, *Physical Review B* **106**, L241106 (2022).

[52] T. Kato, Y. K. Li, K. Nakayama, Z. W. Wang, S. Souma, M. Kitamura, K. Horiba, H.

Kumigashira, T. Takahashi, and T. Sato, Polarity-dependent charge density wave in the kagome superconductor CsV$_3$Sb$_5$, *Physical Review B* **106**, L121112 (2022).

[53] T. Kato, Y. K. Li, M. Liu, K. Nakayama, Z. W. Wang, S. Souma, M. Kitamura, K. Horiba, H. Kumigashira, T. Takahashi *et al.*, Surface-termination-dependent electronic states in kagome superconductors $A$V$_3$Sb$_5$ ($A$=K, Rb, Cs) studied by micro-ARPES, *Physical Review B* **107**, 245143 (2023).

[54] X. Zhou, Y. Li, X. Fan, J. Hao, Y. Dai, Z. Wang, Y. Yao, and H.-H. Wen, Origin of charge density wave in the kagome metal CsV$_3$Sb$_5$ as revealed by optical spectroscopy, *Physical Review B* **104**, L041101 (2021).

[55] N. Shumiya, M. S. Hossain, J. X. Yin, Y. X. Jiang, B. R. Ortiz, H. X. Liu, Y. G. Shi, Q. W. Yin, H. C. Le, S. T. Zhan *et al.*, Intrinsic nature of chiral charge order in the kagome superconductor RbV$_3$Sb$_5$, *Physical Review B* **104**, 035131 (2021).

[56] Z. Wang, Y.-X. Jiang, J.-X. Yin, Y. Li, G.-Y. Wang, H.-L. Huang, S. Shao, J. Liu, P. Zhu, N. Shumiya *et al.*, Electronic nature of chiral charge order in the kagome superconductor CsV$_3$Sb$_5$, *Physical Review B* **104**, 075148 (2021).

[57] Y. K. Li, Q. Li, X. W. Fan, J. J. Liu, Q. Feng, M. Liu, C. L. Wang, J. X. Yin, J. X. Duan, X. Li *et al.*, Tuning the competition between superconductivity and charge order in the kagome superconductor Cs(V$_{1-x}$Nb$_x$)$_3$Sb$_5$, *Physical Review B* **105**, L180507 (2022).

[58] Y. Liu, C. C. Liu, Q. Q. Zhu, L. W. Ji, S. Q. Wu, Y. L. Sun, J. K. Bao, W. H. Jiao, X. F. Xu, Z. Ren *et al.*, Enhancement of superconductivity and suppression of charge-density wave in As-doped CsV$_3$Sb$_5$, *Phys. Rev. Mat.* **6**, 124803 (2022).

[59] H. T. Yang, Z. H. Huang, Y. H. Zhang, Z. Zhao, J. N. Shi, H. L. Luo, L. Zhao, G. J. Qian, H. X. Tan, B. Hu *et al.*, Titanium doped kagome superconductor CsV$_{3-x}$Ti$_x$Sb$_5$ and two distinct phases, *Science Bulletin* **67**, 2176 (2022).

[60] G. F. Ding, H. L. Wo, Y. Q. Gu, Y. M. Gu, and J. Zhao, Effect of chromium doping on superconductivity and charge density wave order in the kagome metal Cs(V$_{1-x}$Cr$_x$)$_3$Sb$_5$, *Physical Review B* **106**, 235151 (2022).

[61] Y. M. Oey, B. R. Ortiz, F. Kaboudvand, J. Frassineti, E. Garcia, R. Cong, S. Sanna, V. F. Mitrovic, R. Seshadri, and S. D. Wilson, Fermi level tuning and double-dome superconductivity in the kagome metal CsV$_3$Sb$_{5-x}$Sn$_x$, *Phys. Rev. Mat.* **6**, L041801 (2022).

[62] M. Q. Liu, T. Han, X. R. Hu, Y. B. Tu, Z. Y. Zhang, M. S. Long, X. Y. Hou, Q. G. Mu, and L. Shan, Evolution of superconductivity and charge density wave through Ta and Mo doping in CsV$_3$Sb$_5$, *Physical Review B* **106**, L140501 (2022).

[63] Y. X. Liu, Y. Wang, Y. Q. Cai, Z. Y. Hao, X. M. Ma, L. Wang, C. Liu, J. Chen, L. Zhou, J. H. Wang *et al.*, Doping evolution of superconductivity, charge order, and band topology in hole-doped topological kagome superconductors Cs(V$_{1-x}$Ti$_x$)$_3$Sb$_5$, *Phys. Rev.*

*Mat.* **7**, 064801 (2023).

[64] Q. Xiao, Q. Z. Li, J. J. Liu, Y. K. Li, W. Xia, X. Q. Zheng, Y. F. Guo, Z. W. Wang, and Y. Y. Peng, Evolution of charge density waves from three-dimensional to quasi-two-dimensional in kagome superconductors Cs(V$_{1-x}$M$_x$)$_3$Sb$_5$ (*M* = Nb, Ta), *Phys. Rev. Mat.* **7**, 074801 (2023).

[65] T. Kato, Y. K. Li, K. Nakayama, Z. W. Wang, S. Souma, F. Matsui, M. Kitamura, K. Horiba, H. Kumigashira, T. Takahashi *et al.*, Fermiology and Origin of *T*$_c$ Enhancement in a Kagome Superconductor Cs (V$_{1-x}$Nb$_x$)$_3$Sb$_5$, *Phys. Rev. Lett.* **129**, 206402 (2022).

[66] K. Nakayama, Y. K. Li, T. Kato, M. Liu, Z. W. Wang, T. Takahashi, Y. G. Yao, and T. Sato, Carrier Injection and Manipulation of Charge-Density Wave in Kagome Superconductor CsV$_3$Sb$_5$, *Physical Review X* **12**, 011001 (2022).

[67] R. Khasanov, D. Das, R. Gupta, C. Mielke, M. Elender, Q. Yin, Z. Tu, C. Gong, H. Lei, E. T. Ritz *et al.*, Time-reversal symmetry broken by charge order in CsV$_3$Sb$_5$, *Physical Review Research* **4**, 023244 (2022).

[68] H. Li, D. Oh, M. Kang, H. Zhao, B. R. Ortiz, Y. Oey, S. Fang, Z. Ren, C. Jozwiak, A. Bostwick *et al.*, Small Fermi pockets intertwined with charge stripes and pair density wave order in a kagome superconductor, *Physical Review X* **13**, 031030 (2023).

[69] H. X. Li, G. Fabbris, A. H. Said, J. P. Sun, Y. X. Jiang, J. X. Yin, Y. Y. Pai, S. Yoon, A. R. Lupini, C. S. Nelson *et al.*, Discovery of conjoined charge density waves in the kagome superconductor CsV$_3$Sb$_5$, *Nat. Commun.* **13**, 6348 (2022).

[70] Q. Xiao, Y. H. Lin, Q. Z. Li, X. Q. Zheng, S. Francoual, C. Plueckthun, W. Xia, Q. Z. Qiu, S. L. Zhang, Y. F. Guo *et al.*, Coexistence of multiple stacking charge density waves in kagome superconductor CsV$_3$Sb$_5$, *Physical Review Research* **5**, L012032 (2023).

[71] L. Kautzsch, B. R. Ortiz, K. Mallayya, J. Plumb, G. Pokharel, J. P. C. Ruff, Z. Islam, E. A. Kim, R. Seshadri, and S. D. Wilson, Structural evolution of the kagome superconductors *A*V$_3$Sb$_5$ (*A* = K, Rb, and Cs) through charge density wave order, *Phys. Rev. Mat.* **7**, 024806 (2023).

[72] Z. C. Jiang, H. Y. Ma, W. Xia, Z. T. Liu, Q. Xiao, Z. H. Liu, Y. C. Yang, J. Y. Ding, Z. Huang, J. Y. Liu *et al.*, Observation of Electronic Nematicity Driven by the Three-Dimensional Charge Density Wave in Kagome Lattice KV$_3$Sb$_5$, *Nano Lett* **23**, 5625 (2023).

[73] S.-L. Yu and J.-X. Li, Chiral superconducting phase and chiral spin-density-wave phase in a Hubbard model on the kagome lattice, *Physical Review B* **85**, 144402 (2012).

[74] W. H. Ko, P. A. Lee, and X. G. Wen, Doped kagome system as exotic superconductor, *Physical Review B* **79**, 214502 (2009).

[75] Y. F. Xie, Y. K. Li, P. Bourges, A. Ivanov, Z. J. Ye, J. X. Yin, M. Z. Hasan, A. Y. Luo,

Y. G. Yao, Z. W. Wang *et al.*, Electron-phonon coupling in the charge density wave state of CsV$_3$Sb$_5$, *Physical Review B* **105**, L140501 (2022).

[76] E. Uykur, B. R. Ortiz, S. D. Wilson, M. Dressel, and A. A. Tsirlin, Optical detection of the density-wave instability in the kagome metal KV$_3$Sb$_5$, *Npj Quantum Materials* **7**, 16 (2022).

[77] G. He, L. Peis, E. F. Cuddy, Z. Zhao, D. Li, Y. Zhang, R. Stumberger, B. Moritz, H. Yang, H. Gao *et al.*, Anharmonic strong-coupling effects at the origin of the charge density wave in CsV$_3$Sb$_5$, *Nat. Commun.* **15**, 1895 (2024).

[78] H.-S. Xu, Y.-J. Yan, R. Yin, W. Xia, S. Fang, Z. Chen, Y. Li, W. Yang, Y. Guo, and D.-L. Feng, Multiband superconductivity with sign-preserving order parameter in kagome superconductor CsV$_3$Sb$_5$, *Phys. Rev. Lett.* **127**, 187004 (2021).

[79] C. Zhao, L. Wang, W. Xia, Q. Yin, J. Ni, Y. Huang, C. Tu, Z. Tao, Z. Tu, and C. Gong, Nodal superconductivity and superconducting domes in the topological Kagome metal CsV$_3$Sb$_5$, *arXiv:2102.08356* (2021).

[80] C. Mu, Q. Yin, Z. Tu, C. Gong, H. Lei, Z. Li, and J. Luo, S-wave superconductivity in kagome metal CsV$_3$Sb$_5$ revealed by $^{121/123}$Sb NQR and $^{51}$V NMR measurements, *Chinese Physics Letters* **38**, 077402 (2021).

[81] W. Y. Duan, Z. Y. Nie, S. S. Luo, F. H. Yu, B. R. Ortiz, L. C. Yin, H. Su, F. Du, A. Wang, Y. Chen *et al.*, Nodeless superconductivity in the kagome metal CsV$_3$Sb$_5$, *Science China Physics, Mechanics & Astronomy* **64**, 107462 (2021).

[82] R. Gupta, D. Das, C. H. Mielke, Z. Guguchia, T. Shiroka, C. Baines, M. Bartkowiak, H. Luetkens, R. Khasanov, Q. W. Yin *et al.*, Microscopic evidence for anisotropic multigap superconductivity in the CsV$_3$Sb$_5$ kagome superconductor, *npj Quantum Materials* **7**, 49 (2022).

[83] Z. Guguchia, C. Mielke, D. Das, R. Gupta, J. X. Yin, H. Liu, Q. Yin, M. H. Christensen, Z. Tu, C. Gong *et al.*, Tunable unconventional kagome superconductivity in charge ordered RbV$_3$Sb$_5$ and KV$_3$Sb$_5$, *Nat. Commun.* **14**, 153 (2023).

[84] Y. Gu, Y. Zhang, X. Feng, K. Jiang, and J. Hu, Gapless excitations inside the fully gapped kagome superconductors $A$V$_3$Sb$_5$, *Physical Review B* **105**, L100502 (2022).

[85] J. A. Sobota, Y. He, and Z.-X. Shen, Angle-resolved photoemission studies of quantum materials, *Reviews of Modern Physics* **93**, 025006 (2021).

[86] K. Okazaki, Y. Ota, Y. Kotani, W. Malaeb, Y. Ishida, T. Shimojima, T. Kiss, S. Watanabe, C. T. Chen, K. Kihou *et al.*, Octet-line node structure of superconducting order parameter in KFe$_2$As$_2$, *Science* **337**, 1314 (2012).

[87] R. Lou, A. Fedorov, Q. Yin, A. Kuibarov, Z. Tu, C. Gong, E. F. Schwier, B. Büchner, H. Lei, and S. Borisenko, Charge-Density-Wave-Induced Peak-Dip-Hump Structure and

the Multiband Superconductivity in a Kagome Superconductor CsV$_3$Sb$_5$, *Phys. Rev. Lett.* **128**, 036402 (2022).

[88] R. Gupta, D. Das, C. Mielke, E. T. Ritz, F. Hotz, Q. Yin, Z. Tu, C. Gong, H. Lei, T. Birol *et al.*, Two types of charge order with distinct interplay with superconductivity in the kagome material CsV$_3$Sb$_5$, *Communications Physics* **5**, 232 (2022).

[89] S. C. Holbæk, M. H. Christensen, A. Kreisel, and B. M. Andersen, Unconventional superconductivity protected from disorder on the kagome lattice, *Physical Review B* **108**, 144508 (2023).

[90] X. X. Wu, D. Chakraborty, A. P. Schnyder, and A. Greco, Crossover between electron-electron and electron-phonon mediated pairing on the kagome lattice, *Physical Review B* **109**, 014517 (2024).